\documentclass[useAMS,usenatbib,twocolumn]{mn2e}
\usepackage{graphicx}
\usepackage[english]{babel}
\usepackage{multirow}
\usepackage{lscape}
\usepackage{longtable}
\usepackage{natbib}
\usepackage[usenames,dvipsnames]{color}
\usepackage{float}
\usepackage{url}
\usepackage[T1]{fontenc}
\usepackage{xspace}
\usepackage{array}
\usepackage{amssymb}
\usepackage{cleveref}

\def\LaTeX{L\kern-.36em\raise.3ex\hbox{a}\kern-.15em
T\kern-.1667em\lower.7ex\hbox{E}\kern-.125emX}

\newcommand{\OII}{$\left[\mathrm{O\,\textrm{\textsc{ii}}}\right]$\xspace}
\newcommand{\OIIll}{$\left[\mathrm{O\,\textrm{\textsc{ii}}}\right]\,(\lambda\lambda 3726,3729)$\xspace}
\newcommand{\OIII}{$\left[\mathrm{O\,\textrm{\textsc{iii}}}\right]$\xspace}

\newcommand{\OIIIa}{$\left[\mathrm{O\,\textrm{\textsc{iii}}}\right]\,(\lambda 4959)$\xspace}
\newcommand{\OIIIb}{$\left[\mathrm{O\,\textrm{\textsc{iii}}}\right]\,(\lambda 5007)$\xspace}
\newcommand{\NeIII}{$\left[\mathrm{Ne\,\textrm{\textsc{iii}}}\right]$\xspace}
\newcommand{\NII}{$\left[\mathrm{N\,\textrm{\textsc{ii}}}\right]$\xspace}

\newcommand{\SII}{$\left[\mathrm{S\,\textrm{\textsc{ii}}}\right]$\xspace}

\newcommand{\Ha}{H${\alpha}$\xspace}
\newcommand{\Hb}{H${\beta}$\xspace}
\newcommand{\Hbl}{H${\beta}\,(\lambda 4861)$\xspace}

\newcommand{\ulum}[0]{${\rm erg \cdot s^{-1}}$\xspace}
\newcommand{\uflux}[0]{${\rm erg \cdot s^{-1} \cdot cm^{-2}}$\xspace}
\newcommand{\ufluxA}[0]{${\rm erg \cdot s^{-1} \cdot cm^{-2} \cdot \mathring{A}^{-1}}$\xspace}

\begin{document}

\title[\OII, \Hb, \OIII LFs]{The evolution of the \OII, \Hb and \OIII emission-line luminosity functions over the last nine billions years} 
\author[J. Comparat et al.]{
 \parbox{\textwidth}{
Johan Comparat$^{1,2}$\thanks{j.comparat@csic.es}\thanks{Severo Ochoa IFT Fellow}, Guangtun Zhu$^{3}$, 
Violeta Gonzalez-Perez$^{4,5}$, Peder Norberg$^{5}$, Jeffrey Newman$^{6}$, Laurence Tresse$^{8}$, Johan Richard$^{8}$, Gustavo Yepes$^{2}$, Jean-Paul Kneib$^{9}$, Anand Raichoor$^{10}$, Francisco Prada$^{1}$, Claudia Maraston$^{4}$, Christophe Y\`eche$^{10}$, Timoth\'ee Delubac$^{9}$, Eric Jullo$^{7}$.}
\vspace*{6pt} \\
$^1$Instituto de F\'{\i}sica Te\'orica UAM/CSIC, 28049 Madrid, Spain\\
$^2$Departamento de F\'{\i}sica Te\'orica, Universidad Aut\'onoma de Madrid, 28049 Madrid, Spain\\
$^3$Department of Physics \& Astronomy, Johns Hopkins University, 3400 N. Charles Street, Baltimore, MD 21218, USA\\
$^4$Institute of Cosmology and Gravitation, University of Portsmouth, Portsmouth, PO1 3FX, UK\\
$^5$Institute for Computational Cosmology and Centre for Extragalactic Astronomy, Department of Physics, University of Durham,\\ South Road, Durham, DH1 3LE, U.K. \\
$^6$Department of Physics and Astronomy and PITT PACC, University of Pittsburgh, Pittsburgh, PA 15260, USA\\
$^7$Aix Marseille Universit\'e, CNRS, LAM (Laboratoire d'Astrophysique de Marseille) UMR 7326, F-13388, Marseille, France\\
$^8$Univ Lyon, Univ Lyon 1, Ens de Lyon, CNRS, Centre de Recherche Astrophysique de Lyon UMR5574, F-69230, Saint-Genis-Laval, France\\
$^9$Laboratoire d'Astrophysique, Ecole Polytechnique F\'ed\'erale de Lausanne (EPFL), Observatoire de Sauverny, CH-1290 Versoix, Switzerland\\
$^{10}$CEA, Centre de Saclay, IRFU/SPP, F-91191 Gif-sur-Yvette, France}
\maketitle
\label{firstpage}

\begin{abstract}
Emission line galaxies are one of the main tracers of the large-scale structure to be targeted by the next-generation dark energy surveys. 
To provide a better understanding of the properties and statistics of these galaxies, 
we have collected spectroscopic data from the VVDS and DEEP2 deep surveys
and estimated the galaxy luminosity functions (LFs) of three distinct emission lines, \OIIll ($0.5 < z < 1.3$), \Hbl ($0.3 < z < 0.8$) and \OIIIb ($0.3 < z < 0.8$). 
Our measurements are based on 35,639 emission line galaxies and cover a volume of $\sim10^7$Mpc$^3$. We present the first measurement of the \Hb LF at these redshifts. 
We have also compiled LFs from the literature that were based on independent data or covered different redshift ranges, 
and we fit the entire set over the whole redshift range with analytic Schechter and Saunders models, assuming a natural redshift dependence of the parameters.
We find that the characteristic luminosity ($L_*$) and density ($\phi_*$) of all LFs increase with redshift. Using the Schechter model over the redshift ranges considered, we find that, for \OII emitters, the characteristic luminosity $L_*(z=0.5)=3.2\times10^{41}$ \ulum increases by a factor of $2.7 \pm 0.2$ from z=0.5 to 1.3; for \Hb emitters $L_*(z=0.3)=1.3\times10^{41}$ \ulum increases by a factor of $2.0 \pm 0.2$ from z=0.3 to 0.8; and for \OIII emitters $L_*(z=0.3)=7.3\times10^{41}$ \ulum increases by a factor of $3.5 \pm 0.4$ from z=0.3 to 0.8.
\end{abstract}

\begin{keywords}
catalogues - surveys - galaxies: abundances - galaxies: evolution - galaxies: general - cosmology: observations.
\end{keywords}


\section{Introduction}
\label{sec:introduction}
Recent precise observations of the cosmic microwave background, the surpernovae, and the Cepheids at the two ends of the observable Universe ($z\sim1100$ and $z\sim0$) 
has led to the $\Lambda$CDM concordance model \citep[e.g.,][]{Freedman_2010,Planck_2014,Suzuki_2012}. 
It successfully describes the evolution of the homogeneous Universe, though requires exotic elements such as dark matter and dark energy \citep[e.g.,][]{Frieman2008}. 

For a better understanding of the puzzling dark ingredients, it is important 
to investigate what happened between the two ends and measure directly the expansion history of the Universe.

Observations now revolve around the inhomogeneous Universe, in particular, by using the baryon acoustic oscillation (BAO) as a standard ruler \citep[e.g., ][]{cole_2005,eisenstein_2005}. The Baryonic Oscillation Spectroscopic Survey (BOSS) experiment demonstrated the ability of BAO measurements to provide a standard ruler to the percent level at redshift 0.5 \citep{Anderson2014}.
Though the BAO is known to be largely free of systematic errors \citep[e.g.,][]{vargas_2014,ross_2015}, precise measurement requires surveys of a large number of sources over a large cosmic volume. 
The current and future dark-energy surveys aim to efficiently sample tens of millions of faint galaxies/quasars at redshift $0.5\lesssim z \lesssim 2.3$ over the entire observable sky to measure BAO at the percent level ({\it e.g.} eBOSS\footnote{\url{http://sdss.org/}}, DESI \footnote{\url{http://desi.lbl.gov/}}, PFS\footnote{\url{http://sumire.ipmu.jp/}}, 4MOST\footnote{\url{https://www.4most.eu/}}, \textit{EUCLID}\footnote{\url{http://sci.esa.int/euclid/}}).

The eBOSS survey \citep{dawson_2015} in the SDSS-IV (Blanton et al., in preparation), started in Fall 2014 and is currently mapping the large-scale structure at redshift $z>0.6$ with four different tracers: luminous red galaxies \citep[LRGs,][]{Prakash2015}, emission line galaxies \citep[ELGs,][]{comparat_2015_elg}, QSOs, and Lyman $\alpha$ absorption \citep[][]{Myers2015,2016A&A...589C...2P}. It will provide the first density map covering the redshift range $0.6<z<2.5$ over a large portion of the sky (7,500 deg$^2$ of LRGs and QSOs, 1,500 deg$^2$ of ELGs).

To achieve these goals with a 2.5-metre telescope, eBOSS requires efficient methods to pre-select targets using both magnitude and colour cuts. For ELGs, there is an additional source of complication: their redshifts will be mainly determined by emission lines, and it is therefore important to characterize the sampling efficiency of ELGs with broad-band magnitude and colour selections \citep[e.g,][]{comparat_2015_elg,Raichoor2016}. Studies have shown that the future ELG surveys will sample the \OII or the \Ha emitters in an incomplete manner \citep{comparat_2015_OII,Tonegawa2015}. A more precise quantification of this incompleteness is critical for a more detailed understanding of the selection effects and to enable more precise cosmological analysis.

In the redshift range targeted by eBOSS (0.6$<z<$1.2 for ELGs), the strongest emission-lines that can be observed in the optical are the oxygen \OIIll, \OIIIb lines and the hydrogen \Hbl Balmer line. All other lines are either outside of the optical window at these redshifts, or an order of magnitude weaker and therefore will not drive the completeness of the ELG survey. Of these three line luminosity functions (LFs), the \OII LF is the most well-known. Its LF has been measured from redshift 0 to 4.7 \citep[e.g.,][]{Ly_2007,Zhu_2009,Gilbank_2010,Bayliss_2011,Sobral_2012,comparat_2015_OII,Khostovan_01102015}, mostly due to interests in using \OII as an empirical star formation rate (SFR) indicator \citep[see][]{Kennicutt_1998,2004AJ....127.2002K,2006ApJ...642..775M}. The \OIII and \Hb LFs have been studied between redshift 0 and 3.3 mostly with narrow-band imaging and are usually measured together (including \OIII$\lambda4959$) as a single LF due to the low spectral resolution \citep[e.g.,][]{Ly_2007, Drake_2013,Khostovan_01102015, Sobral_11082015}. Over the redshift range we are interested in, we did not find a measurement of the \Hbl LF alone (without \OIII).

We present here independent measurements of the \OIIll (total luminosity in the two lines), \Hbl and \OIIIb (\OIIIa is not included) LFs, by taking advantage of the large spectroscopic data sets made available by recent deep surveys,
the VIsible MultiObject Spectrograph Very Large Telescope Deep Survey\footnote{\url{http://cesam.lam.fr/vvdspub/}} (VVDS)
and the DEEP2 survey\footnote{\url{http://deep.ps.uci.edu/}} \citep[][respectively]{lefevre2013,Newman_2013}. 
In addition, we fit the observed LF with two models, the Schechter function \citep{Schechter1976} and the Saunders function \citep{Saunders_1990} over the entire redshift range.

Throughout the paper, we use AB magnitudes \citep{1982STIN...8311000O} and provide measurements in a flat $\Lambda$CDM cosmology \citep[$h=0.677$, $\Omega_\textrm{m0}=0.307$;][]{Planck_2014}.

Spectra, catalogues, LFs, code and fitting functions are publicly available through the skies and universes data base.\footnote{\url{http://projects.ift.uam-csic.es/skies-universes/}}


\section{Data}
\label{sec:DATA}

\begin{figure*}
\begin{center}
\includegraphics[width=17cm]{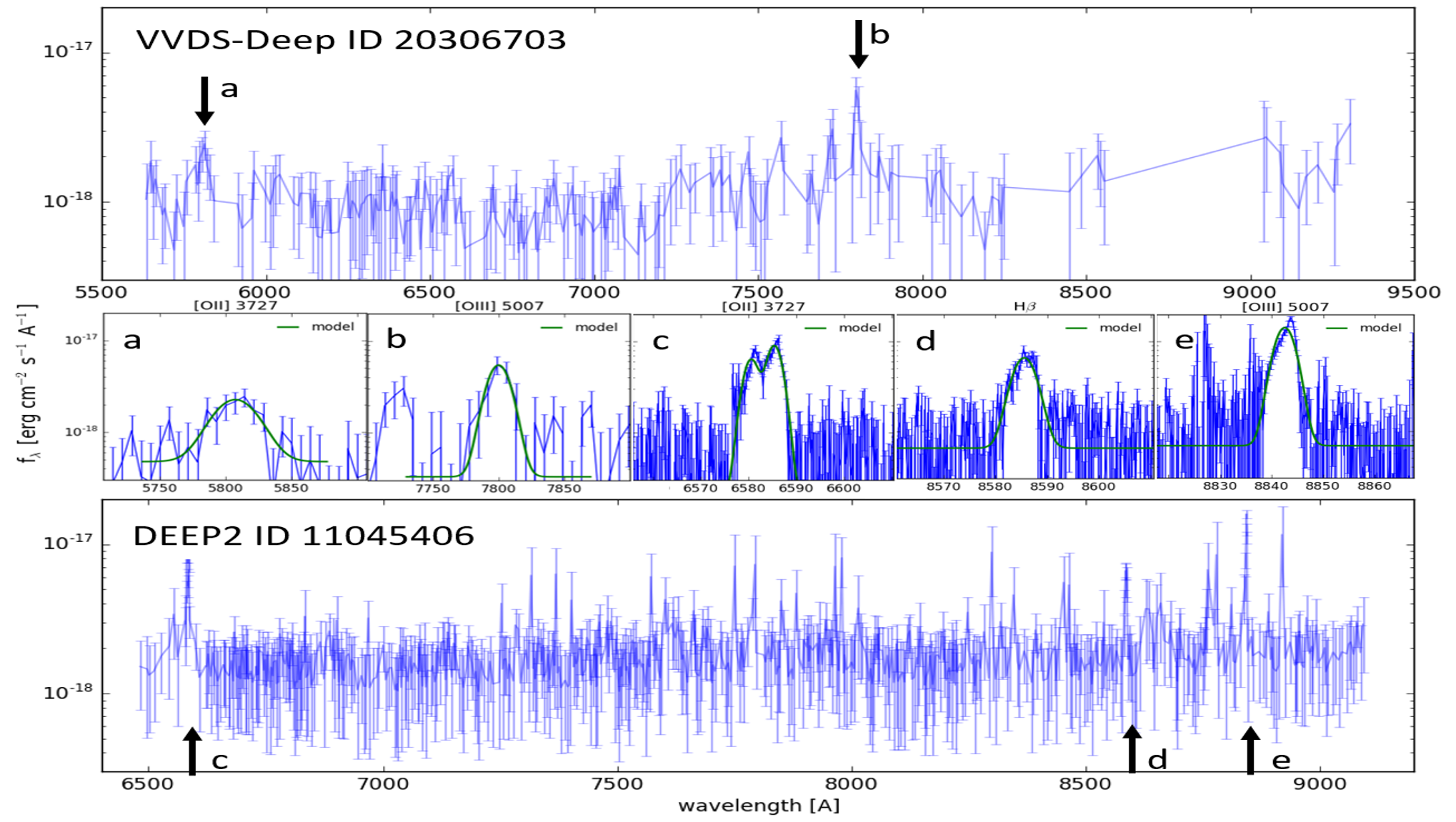}
\caption{Typical VVDS-Deep (top row) and DEEP2 (bottom row) spectra processed with the emission line pipeline. The observed spectrum is shown in blue error bars. We show only points with an SNR larger than 1.5 (DEEP2) and 1.2 (VVDS), otherwise the plot would be dominated by noise. For the DEEP2 spectrum we only show every five points. In the middle row of panels, we show a zoom on the emission lines found in each spectra (no filtering, all data points are shown) and their best-fitting model (green solid line). The VVDS-Deep galaxy (ID 20306703) has a magnitude $i$=23.9 and a redshift $z=$0.55. The DEEP2 galaxy spectrum (ID 11045406) has a magnitude $R=23.7$ and a redshift $z=$0.76. The DEEP2 spectrum is shown after applying flux calibration. The VVDS spectrum shown is corrected from aperture, the emission line fits, performed before correction are plotted along with the uncorrected spectrum.}
\label{fig:spec:example}
\end{center}
\end{figure*}

From the VVDS and DEEP2 survey data bases, we collected spectra of 69,529 unique galaxies with a reliable redshift $z>0.1$. 
Among them, 35,639 have at least one emission line with signal to noise ratio (SNR) greater than 5. 
We present the summary of the data sets in Table~\ref{table:data:one} and describe each data set in Sections \ref{subsec:vvds} and \ref{subsec:deep2}. 
We detail the procedure of emission line fitting in Section \ref{subsec:elfit} and give the line flux completeness in Section \ref{subsec:fluxlimit}.

\begin{table*}
\caption{The spectroscopic data. The total sample contains 69,529 unique galaxy spectra and among them 35,639 with emission-lines. The effective (non-masked) area is given in the second column. We give the magnitude cut applied in each survey in the column `mag. limit' and provide the bands eventually used for a colour selection in the column `colour selection'. The $\mathcal{R}$ and $\lambda$ columns give the resolution at the median wavelength and the wavelength range covered by the spectrographs.
$N_{\rm total}$ is the number of galaxies with an estimated redshift at z$>0.1$ (VVDS: Zflags$>=1$, DEEP2: ZQUALITY$>=2$). 
$N_{\rm lines}$ is the subset of galaxies for which at least one of the \OII or \Hb or \OIII lines has a ${\rm SNR}>5$.
All 35,639 with emission-lines have redshift quality flags VVDS: Zflags$>=2$, DEEP2: ZQUALITY$>=3$.}
\begin{center}
\begin{tabular}{c c c c c c c c c c c }
\hline \hline
Survey	 & Area (deg$^2$) & Mag. limit & colour selection & $\mathcal{R}$ & $\lambda$ (nm) & $N_{\rm total}$ & $N_{\rm lines}$ (\OII; \Hb; \OIII) \\ \hline
VVDS-Deep & 0.6 			& $i_{\rm AB}<24$ 		& No 	 	& 230 		& 550 935 & 10,123 & 3,833 (2,853; 472; 1,226) \\
VVDS-Wide & 5.8 			& $i_{\rm AB}<22.5$ 	& No 		& 230 		& 550 935 	& 23,993 	& 5,909 (3,652; 911; 2,334) \\ 
DEEP2 Field 1 (EGS)	& 0.5	&		$R_{\rm AB}<24.1$	& No 	& 6000 	& 640 910 	& 12,263 & 8,374 (4,444; 3,202; 4,144) \\ 
DEEP2 Fields 2, 3, 4 & 3.0	&		$R_{\rm AB}<24.1$	& BRI 	& 6000 	& 640 910 	& 23,140 & 17,523 (15,358; 3,800; 3,364) \\ \hline
\end{tabular}
\end{center}
\label{table:data:one}
\end{table*}%

\subsection{VVDS}
\label{subsec:vvds}
VVDS \citep{lefevre2013} was conducted with the visible wide field imager and multi-object spectrograph (VIMOS) mounted on the Nasmyth focus B of UT3 Melipal of 
the European Southern Observatory Very Large Telescope located in Chile \citep{2003SPIE.4841.1670L}. 
VVDS is a magnitude-limited survey and includes a Wide component and a Deep component. 
The magnitude limits are $i=22.5$ and $24.0$ and the effective areas covered are 5.8 and 0.6 ${\rm deg}^2$ for the Wide and the Deep, respectively.
The targets were chosen using the $i$ magnitude from CFHT observations \citep{McCracken2003,Ilbert2005,Cucciati2012}. 
The VVDS collaboration provides the slit-extracted 1D-spectra and the redshift catalogue based on visual inspection of the spectra. 
The spectral resolution $\mathcal{R}$ is about $\sim230$ and the wavelength coverage is from $550$ to $935\,$nm. 
The exposure time ranges from 0.75h for the Wide to 4.5h for the Deep.
Further information on VVDS may be found in \citet{lefevre2013}. 
We obtain 5,909 and 3,833 ELGs with at least one of the three \OII, \Hb and \OIII lines with ${\rm SNR}>5$ in the VVDS-Wide and VVDS-Deep fields, respectively.

\subsection{DEEP2}
\label{subsec:deep2}
DEEP2 \citep{Newman_2013} was conducted with DEep Imaging Multi-Object Spectrograph \citep[DEIMOS;][]{2003SPIE.4841.1657F} mounted on the Nasmyth focus of the Keck II telescope. 
In DEEP2 Fields 2, 3, 4, the survey is complete down to $R=24.1$ at redshift $z>0.7$, 
with the desired redshift range achieved with a pre-selection on the $B-R$ and $R-I$ plane,
while Field 1 (the Extended Groth Strip, EGS) did not include the colour pre-selection and is complete over the entire redshift range ($0<z\lesssim1.4$).
DEEP2 is a complete galaxy survey for redshifts $z>0.7$ and magnitude $R\leq24.1$. 
The DEEP2 collaboration released redshift catalogues and 1D slit-extracted spectra. 
The spectral resolution $\mathcal{R}$ is about $6000$ and the wavelength coverage is from $640$ to $910\,$nm. 
Further information on DEEP2 may be found in \citet{Newman_2013}. 
We obtain 25,897 ELGs from the DEEP2 data.

\subsection{From spectra to emission line catalogues}
\label{subsec:elfit}

We first construct catalogues of emission lines based on the spectra and redshift catalogues provided by the aforementioned surveys.
We use a single routine to fit the flux of emission lines in the galaxy spectra across the different surveys, 
inspired by the SDSS pipeline \citep{2012AJ....144..144B,2013MNRAS.431.1383T}. 
As the resolution varies from one survey to another, $\mathcal{R}\sim230$ for VVDS and $\sim6000$ for DEEP2, 
we cannot extract the same level of information from all the spectra 
and our code includes free parameters to treat properly the different resolutions.
In this paper, although we focus on the \OIIll (sum of the two lines), \Hbl and \OIIIb (note that \OIIIa is not included) lines, we extend the search to a series of lines for future analysis.
Table \ref{table:EMlineList} summarizes the list of emission lines we search for in each spectrum. 
To avoid the contamination from the strong sky lines in the red, we only search for lines up to $\lambda=9000\,$\AA.

\begin{table}
\caption{The list of emission lines searched for in the spectra, with wavelengths collected from \citet{pyneb2015}. 
We split the lines into collisional and recombination groups. 
The last column gives the redshift where the line's observer-frame wavelength is $9000\,$\AA, the maximum wavelength considered in the line identification We quote wavelengths in the air (for VVDS, DEEP2, VIPERS) and in the vacuum (for SDSS) because surveys provide spectra wavelengths in one of the conventions but not both.}
\begin{center}
\begin{tabular}{ c c c c}
\hline\hline
Line & \multicolumn{2}{c}{Rest-frame wavelength [\AA] } & $z$ at \\
 & vacuum & air & 9000\,\AA \\ \hline
 \multicolumn{3}{c}{Collisional lines} & \\ \hline
\OII & 3727.092 & 3726.032 & 1.41 \\ 
\OII & 3729.875 & 3728.815 & 1.41 \\ 
\NeIII & 3869.861 & 3868.764 &1.32 \\ 
\OIII & 4364.436 & 4363.209 &1.06 \\ 
\OIII & 4960.295 & 4958.910 & 0.81 \\ 
\OIII & 5008.240 & 5006.842 & 0.79 \\ 
\NII & 6549.861 & 6548.049 & 0.37 \\ 
\NII & 6585.273 & 6583.451 & 0.36 \\ 
\SII & 6718.295 & 6716.437 & 0.34\\ 
\SII & 6732.674 & 6730.812 & 0.33\\ 

\hline
 \multicolumn{3}{c}{Recombination lines} \\ \hline
H $\epsilon$ & 3971.202 & 3970.079 & 1.26\\ 
H $\delta$ & 4102.899 & 4101.741 & 1.19\\ 
H $\gamma$ & 4341.691 & 4340.470 &1.07 \\ 
H $\beta$ & 4862.691 & 4861.332 & 0.85\\ 
H $\alpha$ & 6564.632 & 6562.816 &0.37 \\ 
\hline
\end{tabular}
\end{center}
\label{table:EMlineList}
\end{table}%

\medskip

Our emission-line measurement pipeline includes the following steps.
\begin{enumerate}
\item Flux-calibrate the DEEP2 spectra. 
As the spectra given by the DEEP2 collaboration are not flux-calibrated, 
we perform the calibration ourselves with the broad-band photometry.
The calibration procedures includes the following items.
\begin{enumerate}
\item The correction of the quantum efficiency of the detector chips.
\item The correction of the $A$ and $B$ telluric absorption bands.
\item The flux calibration using the $R$ and $I$ total photometry, assuming that the shape of the observed spectrum within the slit
is representative of that of the spectral energy distribution (SED) emitted by the whole galaxy. 
\end{enumerate}
\item The correction of aperture effects of the VIMOS spectra. The slit spectra from VVDS are already calibrated for spectrophotometry. 
We apply additional aperture corrections to convert the flux within the slit to the total flux of the whole galaxy:
\begin{enumerate}
\item Integrate the observed spectrum over the $i$-band filter (which is the selection band of the survey),
\begin{equation}
m_{\rm spec}=\frac{\int_{\rm broad-band} d\lambda \; f_\lambda(\lambda) {\rm filter}(\lambda) }{ \int_{\rm broad-band} d\lambda \; {\rm filter}(\lambda) }\ \mathrm{;}
\end{equation}
\item Convert the $m_{\rm spec}$ into AB magnitude $m_{\rm spec,\,AB}$ and compare $m_{\rm spec,\,AB}$ with the total magnitude taken from the targeting photometry, 
assuming the shape of the observed spectrum within the slit represents that of the total SED of the whole galaxy.
\end{enumerate}
\item Determine the observer frame wavelength of the emission lines listed, denoted $\bar{\lambda}$, listed in Table~\ref{table:EMlineList} using the best redshifts from the catalogues.
\item Estimate the continuum flux density $C_{\bar{\lambda}}$ (in \ufluxA), using the median of the observed spectrum in a wide (dependent on the resolution) spectral band on the left-hand side or on the right-hand side of the expected line location. 
\item Compare the mean flux density in the two closest pixels to the expected line position with the flux density of the continuum ($C_{\bar{\lambda}}$). If this ratio is greater than one, we fit a line model. 
\item Fit each line with a Gaussian model:
 \begin{equation}
 f^{\rm G}_\lambda(\lambda,\bar{\lambda} ,\sigma,F,C_{\bar{\lambda}}) = C_{\bar{\lambda}} + F \, \frac{{\rm e}^{-(\lambda-\bar{\lambda})^2/(2 \sigma^2)} }{\sigma\sqrt{2{\rm\pi}}}.
 \end{equation}
The free parameters are the total flux, $F$, in \uflux, the width, $\sigma$, in \AA.
\item Fit the \OII doublet with a double-Gaussian profile. 
Because the high resolution of the DEIMOS spectrograph, we are able to resolve the \OII doublet. 
The double-Gaussian profile is given by
\begin{equation}
f_\lambda(\lambda ,\sigma,F,y,C_{\bar{\lambda}}) = C_{\bar{\lambda}} + \frac{F}{\sigma\sqrt{2{\rm\pi}}} \left[\frac{ (1-y)}{{\rm e}^{\frac{(\lambda-3726)^2}{2 \sigma^2}}} + \frac{y}{ e^{\frac{(\lambda-3729)^2}{2 \sigma^2}} }\right]
\end{equation}
The line flux ratio is thus given by $F_{3729} / F_{3726} = y/(1-y)$.
For the data observed with VIMOS at resolution $\mathcal{R}\sim230$, 
we cannot fit for the $y$ parameter and we fix it to be 0.58, 
its mean expected value \citep{Pradhan2006}. If we fit a single Gaussian, the fits converge as well. But the width of the line will encompass both lines and it is less convenient to compare it to the width of other lines. 
\end{enumerate}

Fig. \ref{fig:spec:example} shows two examples of spectra (one from VVDS-Deep and one from DEEP2) as well as the model fitted to the emission lines. The differences due to the discrepant resolution are clear.

When the fitting fails (e.g., due to masked pixels, high sky residuals), 
we output in the catalogue the estimates of the continuum $C_{\bar{\lambda}}$ and the flux density in the two pixels nearest to the expected line position. 

\subsubsection*{Limitations}

We do not correct for the Balmer absorption intrinsic to the underlying stellar continuum, which requires high SNR. In the DEEP2 spectra out of 7,002 with an \Hb fit, we found two galaxies with a continuum SNR above $10$ around the \Hb lines and a negative flux fitted. This is a small fraction of galaxies where we could be biased in the estimation of the line flux by not accounting for the absorption. 

In typical star-forming galaxies, star formation and thus line emission can be more extended than stellar continuum \citep[e.g.][]{ForsterSchreiber2011,Nelson2013}. As a consequence, the use of slit or aperture with a limited size may introduce a bias in the total flux measurements. \citet{Nelson2013} found that, on average, the size of \Ha-emitting region is 1.3 times larger than the $R$-band continuum size for strongly star-forming galaxies at $z\sim1$. The effective size of star-forming galaxies is about 5 10 kpc (in diameter) at $z\sim1$ \citep[e.g.][]{Colbert2013,Nelson2013,Wuyts2013}, and we expect that the arcsec-wide slit used in DEEP2 or VVDS encloses most, if not all, of the line emission. In addition, the DEEP2 team has compared the spatial profiles along the slit for line emission and stellar continuum and no difference was detected (DEEP2 team, private communication).

\subsection{Emission line flux limit}
\label{subsec:fluxlimit}
As we are interested in the statistics of emission line fluxes, we need to understand the completeness of each line measurement, 
i.e., whether or not we can detect the line of interest with an SNR of 5 level at the expected wavelength. 
Ideally, we could determine the line completeness as a function of wavelength (thus redshift) for every individual spectrum, 
which depends on the observing conditions and the sky-background SED.
For simplicity, however, we choose to use a conservative mean flux limit for each survey. 
We use the exposure time calculators (ETC) from VIMOS and DEIMOS to obtain an estimate of the noise level given the instrumental set up of each survey at 8300\AA\xspace, centre of the $I$ band at the redder end of the spectrum where the noise is higher.
We construct a fake spectrum with two components: a constant continuum (in $f_\lambda$) and a single emission line with a Gaussian profile. The integral of the spectrum sums to the limiting magnitude. We vary the relative importance of the emission line to the continuum to obtain a set of spectra that would be observed at the magnitude limit. We fit the emission line using the noise from the ETC and obtain the SNR as a function of line flux.
We obtain the following mean SNR 5 flux limits: $f_{\rm min}^{\rm DEEP2}=2.7$, $f_{\rm min}^{\rm VVDS\, Deep}=1.9$, and $f_{\rm min}^{\rm VVDS\, Wide}=3.5\times10^{-17}\,$\uflux.
In practice, to be conservative, we only consider the luminosity bins that are brighter than the flux limits at a given redshift when measuring the LF.

\subsection{The emission line catalogues}
In total, we detected at least one of the three \OII, \Hb and \OIII emission lines with ${\rm SNR}>5$ in 35,639 spectra (half the sample). All of them have very secure redshift quality flags VVDS: Zflags$>=2$, DEEP2: ZQUALITY$>=3$.
The catalogues are available via the skies and universes data base: 
\url{http://projects.ift.uam-csic.es/skies-universes/LFmodels/content/catalogues/}, they are named:
\begin{tabular}{l}
\url{zcat.deep2.dr4.v4.fits} \\
\url{VVDS_WIDE_summary.v1fits} \\
\url{VVDS_DEEP_summary.v1.fits} \\
\end{tabular}
We report all the line fitting results in the catalogues with the following column naming convention: \textsc{line element, ionization number, wavelength, quantity}. For example, the \OII flux is given in the column `O2\_3728\_flux' and the error on this quantity is given in the column `O2\_3728\_fluxErr'.

\section{Luminosity functions}
\label{sec:LFs}

\begin{figure*}
\begin{center}
\includegraphics[width=17cm]{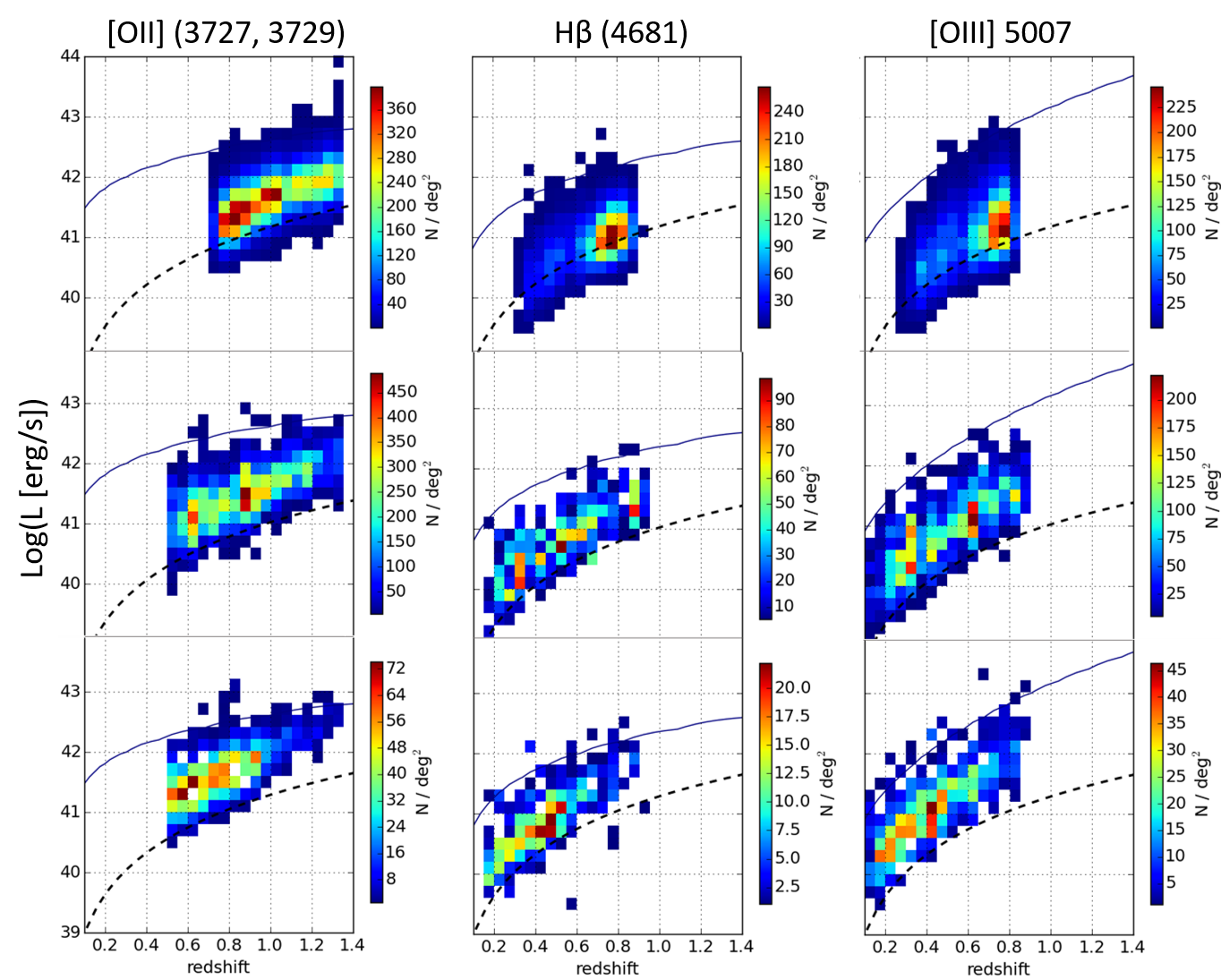}
\caption{\label{fig1:redshift:luminosity}
Number density of emitters per square degree as a function of line luminosity and redshift: DEEP2 (first row of panels), VVDS-Deep (second) and VVDS-Wide (third). 
The dashed lines represent the average $5\sigma$ flux limits given in Section~\ref{subsec:fluxlimit}.
The density of one emitter per square degree predicted by the best-fitting LFs is represented by the solid blue line. Note that the colour bar range vary from panel to panel. It shows how the detection of the lines is limited in redshift, see Table \ref{table:redshift:bins} and text for further details.}
\end{center}
\end{figure*}

The LF, $\Phi(L)$, measures the number of galaxies ($N$) per unit volume ($V$) as a function of luminosity ($L$):
\begin{equation}
dN=\Phi(L) \, dL \, dV.
\end{equation}

In this section, we define the samples (Section \ref{subsec:design}) used for the LF estimate (presented in Section \ref{subsec:LFS}).

\subsection{Samples design}
\label{subsec:design}
As the LFs of the three emission lines, \OII, \Hb and \OIII are spread out in wavelength and redshift, we present here key information on the redshift and luminosity parameter space covered by the data. Fig. \ref{fig1:redshift:luminosity} shows the redshift and luminosity distribution of the emission lines detected in each survey, together with the associated emission line flux limits as estimated in Section \ref{subsec:fluxlimit}.

\subsubsection{Redshift bins}
The redshift range over which lines are detected varies with survey; as shown in Fig. \ref{fig1:redshift:luminosity}.
For each line, within its redshift detection limits, we divide the sample into smaller redshift bins to grasp the evolution of the LF.
We stop at 9000\AA\xspace
to avoid the hydroxyl forest from the atmosphere.
We present the redshift bins in Table~\ref{table:redshift:bins} for the lines and surveys considered. 

In deep pencil beam surveys, the volume covered by the low redshift data is very small, e.g., 
of order of 10$^4$ Mpc$^3$ for $z<0.18$ for a field of view of few square degrees. 
The sample variance of any galaxy population is therefore large and thus we do not consider galaxies at $z<0.18$ in this study.

\begin{table}
\caption{\label{table:redshift:bins}Redshift bins used in the analysis, x means an LF was measured.}
\begin{center}
\begin{tabular}{ cccccc}
\hline\hline
\multicolumn{2}{c}{Redshift} & \multicolumn{3}{c}{Line} \\
min & max & \OII & \Hb & \OIII \\ \hline
\multicolumn{5}{c}{VVDS} \\ \hline
0.18 	& 0.41 	& & x & x \\
0.41 	& 0.65 	& & x & x \\
0.51 	& 0.7 		& x & x & x \\
0.56 	& 0.83 	& x & x & x \\
0.65 	& 0.84 	& x & x & x \\
0.84 	& 1.1 		& x & & \\
1.1 	& 1.3 		& x & & \\ \hline
\multicolumn{5}{c}{DEEP2} \\
\hline
0.33 	& 0.45 & & x & x \\
0.45 	& 0.60 & & x & x \\
0.60 	& 0.70 & & x & x \\
0.7 	& 0.75 & & x & x \\
0.75 	& 0.78 & x & x & x \\
0.78 	& 0.83 & x & x & \\
0.83 	& 1.16 & x & & \\
1.16 	& 1.30 & x & & \\
\hline\hline
\end{tabular}
\end{center}
\end{table}%

\subsubsection{Luminosity limits}
For each LF we compute its luminosity limit, denoted $L^{\rm line}_{\rm min}$, above which the survey is complete. We use a fixed grid of 50 log luminosity bins between 38 and 45 (steps of $\sim0.14$ dex). 
The luminosity limit is constrained by the average $5\sigma$ flux limit $f^{\rm line}_{\rm min}$, as calculated in Section \ref{subsec:fluxlimit}, 
and $L^{\rm line}_{\rm min}$ has to be {\it greater} than $f^{\rm line}_{\rm min} 4 {\rm\pi} d_L^2(z^{\rm line}_{\rm max})$.

Furthermore, for each LF, we measure the line luminosity at which the weighted number counts distribution (see next section for the description of the weights) peaks, $L_{\rm peak}$. 
We find $L_{\rm peak}$ to be always greater than the luminosity limit determined by $L^{\rm line}_{\rm min}$.
To be conservative, we consider the final luminosity limit $L^{\rm line}_{\rm min}$ to be 4 times greater than $L_{\rm peak}$, i.e., we discard the first two bins of luminosity.
We report the luminosity limits for all the lines, redshifts and surveys considered
in Tables \ref{O2:LFs:literature}- ref{OIII:LFs:literature}.

\subsection{LF measurements}
\label{subsec:LFS}

\subsubsection{Sampling rate corrections}
\begin{figure}
\begin{center}
\includegraphics[width=7cm]{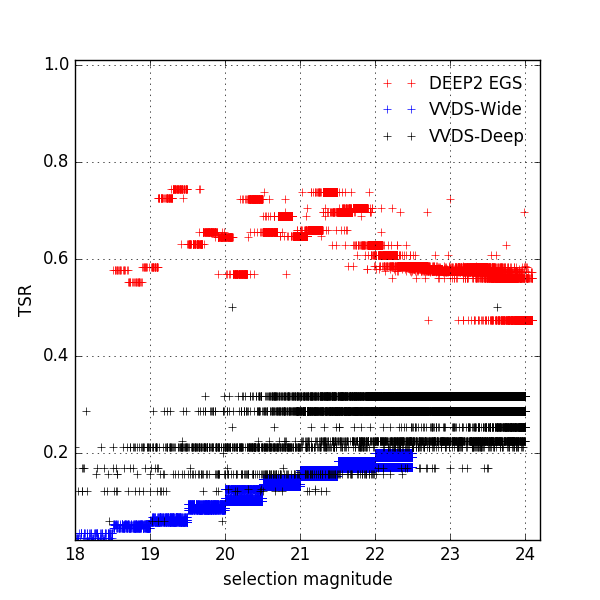}
\includegraphics[width=7cm]{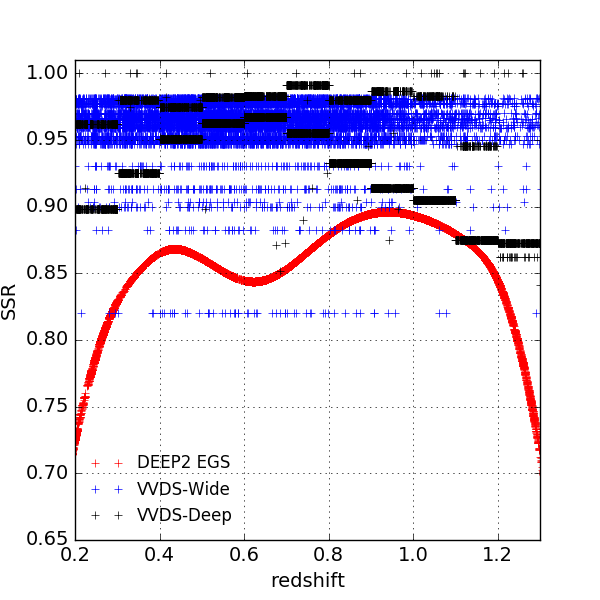}
\caption{TSR as a function of magnitude (top) and SSR as a function of redshift (bottom). We show the values obtained for the following fields: DEEP2 EGS, VVDS-Deep and VVDS-Wide.}
\label{fig:TSR:SSR}
\end{center}
\end{figure}

For each galaxy, we correct the observed densities from the target success rate (TSR) and the spectroscopic success rates (SSR). 
The TSR is the number of targets that were allocated a slit for spectroscopic observation divided by the number of photometric targets down to an apparent magnitude limit. The TSR depends on RA, DEC, magnitudes and for VVDS on the size of the galaxy along the spatial dimension of the slit \citep[see][]{Ilbert2005}. 
The SSR is the fraction of spectroscopic targets for which the redshift was successfully determined. 
The SSR depends on RA, DEC, magnitudes, and redshift.
To compute an LF for a given survey, considering we are interested in the average volume density over the entire (small) area 
covered by the survey, we here choose to ignore the variations of the TSR and SSR as a function of the location 
(though these 
variations would be important for clustering analysis). Furthermore, we bin the data in narrow redshift bins, and therefore, we neglect the variation of TSR with galaxy size. 
VVDS-Deep TSR and SSR are provided through their data base \citep[see][for more details]{Ilbert2005}. For the VVDS-Wide and DEEP2, we compute them as follows. 

We use the targeting photometry to derive the TSR as a function of magnitude. 
We find that DEEP2 has a TSR around 65\% in its field 1 for galaxies with magnitudes $18.5<R<22$ and around 60\% for magnitudes fainter than 22; see Fig. \ref{fig:TSR:SSR} top panel. For DEEP2 Fields 2, 3, 4, we assume the colour pre-selection was efficient and selected galaxies only at $z>0.7$, 
and we use the photometric sample {\it after} the colour pre-selection as the parent sample. 
We obtain a TSR $\sim60\%$.
In these fields, we consider only the redshift range $z>0.7$ in the LF measurements.
The TSR of VVDS-Wide is between 10 and 20\% down to $i<22.5$; see Fig. \ref{fig:TSR:SSR} top panel. 

To estimate the SSR, we use the photometric redshift catalogues from \citet{Ilbert_06,Ilbert_2009} and \citet{Coupon_2009} to complement the spectroscopic redshift catalogues. 
These photometric redshift catalogues cover the same area as the spectroscopic surveys. 
For the SSR, we only account for its dependence on redshift (including colour and magnitude has a negligible effect on our results). 
For each targeted galaxy, if the survey failed to determine the redshift from the spectroscopy, we use the photometric redshift.
We then calculate the SSR as the fraction of galaxies with successful spectroscopic redshift in each redshift bin.
The smallest redshift bin used in the LF analysis has a width of d$z=0.03$ (at $z\sim0.75$) and 
the photometric redshift precision is $\sigma_z/(1+z)\sim0.005(1+0.75)$, which should be sufficient.
In a given redshift bin, the sum of the 1/SSR therefore represents the expected number of (targeted) galaxies.
We determine that DEEP2 has an average SSR$>80\%$ over the redshift range $0.2<z<1.2$, 
VVDS-Deep has an average SSR$>80\%$ at $0.4\lesssim z \lesssim1.2$, and VVDS-Wide has an SSR around $0.95$.

Finally each galaxy with a spectroscopic redshift $z_{\rm spec}$ and an apparent magnitude m receives a weight 1/(TSR(m)*SSR($z_{\rm spec}$)) to account for missed galaxies due to targeting completeness and redshift success rate.

\subsubsection{Measurements and redshift evolution}

We estimate the LF with the samples described above with the non-parametric $V_{\rm max}$ estimator \citep{1968ApJ...151..393S,2011A&ARv..19...41J}. 
We measure in total 18 LFs (\OII: 8, \Hb: 5, and \OIII: 5) in redshift bins between 0.18 and 1.30. 
Sample variance errors are estimated using the jackknife technique.
Using numerous realizations of the COSMOS and SUBARU Deep Field, \citet{Sobral_11082015} derived the uncertainty on the emission line LF parameter estimation as a function of volume. For the volumes considered in this analysis (of the order of $10^6\,{\rm Mpc}^3$ for each LF), we expect errors induced by cosmic variance to be of order of 10 - 20\% (for each LF).

To complement our measurements, we have also gathered from the literature measurement of emission line LFs that cover a volume larger than $\sim10^4$ Mpc$^3$ and span at least half an order of magnitude in luminosity. For the samples that were re-analysed, we only considered the latest version of the measurement. 
The sources, the redshift and luminosity distribution of these measurements of is shown in Table~\ref{O2:LFs:literature} for the \OII doublet, in Table \ref{hbeta:LFs:literature} for the \Hb line and in Table \ref{OIII:LFs:literature} for the \OIII line. 
Under this volume constraint, we have not found any measurement of the observed \Hb emission-line LF alone, i.e. split from \OIIIa and \OIIIb. The combination of all the new and previous \OII, \Hb and \OIII data sets enable the coverage of total volumes of 5, 0.3 and 2$\times10^7$Mpc$^3$, respectively.

We present our measurements\footnote{all the measurements are available here \url{http://projects.ift.uam-csic.es/skies-universes/LFmodels/data}} together with previous measurement from the literature in Figs.~\ref{fig:modelled:LF:O2}-\ref{fig:modelled:LF:O3}.
We observe strong redshift evolution of the LF of all three lines.
Over the redshift range $0\lesssim z \lesssim2.3$, the characteristic luminosity of \OII at a given density (e.g., $10^{-4}\,{\rm Mpc}^{-3} {\rm dex}^{-1}$) 
increases with redshift.
It also shows that for a fixed \OII luminosity e.g. $10^{42}$ \ulum the number density has increased by over a factor of 10 from redshift 0 to 1, consistently with previous studies \citep[e.g.,][]{Zhu_2009,Sobral_11082015}.
For the \Hb and \OIIIb, previous works measured the combined LFs, including \Hb, \OIIIa, and \OIIIb, and found that the number density of systems at a given luminosity (of the three lines combined) increases with redshift \citep[e.g.][]{Ly_2007,Khostovan_01102015,Sobral_11082015}. From our measurements, for the first time, we can disentangle the contributions of each line to this measurement and find that the number density of strong \Hb and \OIII emitters increases with redshift, up to $z\sim0.9$ and $z\sim1.8$, respectively. We measure that the density of \OIIIb emitting galaxies with $L=10^{41}$ \ulum at redshift 0.7 is about twice that of \Hb emitters $\Phi_{\rm [O_{III}]}(L=10^{41}$\ulum$, z=0.7)=2\Phi_{\rm H_\beta}(L=10^{41}$\ulum$, z=0.7)$. At the bright end $L>10^{42}$ \ulum, \OIIIb emitters are 10 times more numerous than \Hb emitters.


\begin{table*}
\caption{Information about the compilation of the \OIIll LF measurements, ordered by mean redshift. Sources. NB=Narrow-Band; S =spectroscopy. The first set of LFs are derived in this paper. Dashes mean the data are the same as given in the line above. Blank spaces mean we lack the information. The combination of these samples covers $\sim5\times10^7$Mpc$^3$.}
\begin{center}
\begin{tabular}{ccc cc c c c c c c c}
\hline \hline
\multicolumn{3}{c}{Redshift} & & Area &N& \multicolumn{2}{c}{ $\log(L$ [\ulum])}& Type & Source\\
mean & min & max & ${\rm\log(\frac{V}{Mpc^3})}$ & (deg$^2$) && min & max \\ \hline
0.81& 0.78& 0.83& 5.99& 2.78& 627 & 41.8& 43.0 & S & DEEP2 \\ 
0.93& 0.83& 1.03& 6.66& 2.78& 4545 & 42.0& 44.0 & S & - \\ 
1.23& 1.16& 1.30& 6.60& 2.78& 926 & 42.4& 44.0 & S & - \\ 

0.61& 0.51& 0.70& 5.77& 0.62& 284 & 41.6& 43.0 & S & VVDS-Deep \\ 
0.74& 0.65& 0.84& 5.88& 0.62& 373 & 41.7& 43.0 & S & - \\ 
0.94& 0.84& 1.10& 6.13& 0.62& 600 & 42.0& 43.0 & S & - \\ 

0.64& 0.51& 0.70& 6.74& 5.79& 504 & 42.8 & 43.1	 & S & VVDS-Wide \\
0.74& 0.65& 0.84& 6.85& 5.79& 411 & 42.8 & 43.1 & S & - \\ \hline

0.10 & 	0.03 &		 0.20 & 	7.20&		275.00	&	43155&	39.7	&	 42.0	&	S & \citet{Gilbank_2010} \\

0.15 & 0.00	&	0.20	& 		4.63	&		0.46&		39	&		 40.0 & 41.5 & S & \citet{Ciardullo2013}\\
0.26 & 0.2	& 	0.32	&		4.90	&	0.46&	70	&		 40.0 & 41.5 & S & -\\
0.38 & 	0.32&		0.45	& 		5.18	&	0.46&		89	&		 40.5 & 42.0 & S & - \\
0.50 & 	0.45&		0.56	& 		5.26	&	0.46&		76	&		 40.5 & 42.0 & S & - \\

 0.17 & 	0.10 & 0.24 & 5.00 & 48.00 & 4450& 40.5 & 42.5 &S & \citet{comparat_2015_OII} \\
0.59 & 0.5 & 0.69 & 5.98 &			 &			4579 & 41.0 & 43.5 &S & - \\
0.78 & 0.69 & 0.88 & 		 6.10 &	 &			 3951 & 41.0 & 43.5 &S & - \\
0.98 & 0.88 & 1.09 & 			6.26 &			 & 1947 & 	41.0 & 43.5 &S & - \\
1.49& 1.34 & 1.65 &			 6.55 &			 & 231 & 42.7 & 44.0 &S & - \\

0.35 & & & 4.38 & 0.38 & 112 & 40.0 & 41.2 &NB & \citet{Drake_2013} \\ %
0.53 & & & 4.63 & 0.38 & 83 & 40.5 & 41.0 &NB & - \\ %
1.19 & 		 &	&	5.31 & 	0.38 & 	981 & 			41.3 & 	42.0 &NB		 & - \\ %
1.64 & & & 5.48 & 0.38 & 27 &	42. & 				43.0 &NB & - \\ %

0.91 & 0.90 & 0.92 & 4.54 & 0.24& 5897 & 41.5 & 42.0 &NB & \citet{Ly_2007} \\ 

2.18 & 2.16& 2.20& 5.54 & 10.00 & 463 & 42.7 & 43.3 & NB and S & \citet{Sobral_11082015} \\
1.47 & 1.45 &1.49 & 5.83 && & 42.0 & 42.6 & NB and S & \citet{Khostovan_01102015} \\
2.25 &2.23 &2.27 &5.79 & & &42.6 & 42.7 & NB and S & - \\
\hline \hline
\end{tabular}
\end{center}
\label{O2:LFs:literature}
\end{table*}

\begin{table*}
\caption{Information about the compilation of the \Hb LF measurements, ordered by mean redshift, only from this paper. The combination of all the samples covers $\sim3\times10^6$Mpc$^3$.}

\begin{center}
\begin{tabular}{ccc cc ccc c c c c}
\hline \hline
\multicolumn{3}{c}{Redshift} & & Area &N& \multicolumn{2}{c}{ $\log(L$ [\ulum])}& Type \\
mean & min & max & ${\rm\log(\frac{V}{Mpc^3})}$& (deg$^2$) && min & max \\ \hline
0.40& 0.33& 0.45& 5.25& 0.60& 269 & 40.2& 42.5 & S \\
0.52& 0.45& 0.60& 5.57& 0.60& 360& 40.6& 42.0 & S \\
0.65& 0.60& 0.70& 5.48& 0.60& 456 & 40.7& 42.3 & S \\
0.75& 0.70& 0.78& 6.12& 2.78& 1142 & 40.9& 42.7& S \\
0.80& 0.78& 0.83& 5.96& 2.78& 739 & 41.0& 42.7& S \\
\hline
\hline
\end{tabular}
\end{center}
\label{hbeta:LFs:literature}
\end{table*}

\begin{table*}
\caption{Information about the compilation of the \OIIIb (\OIIIa is not included) LF measurements, ordered by mean redshift. Sources: NB=Narrow-Band; S =spectroscopy. The combination of these samples covers $\sim2\times10^7$Mpc$^3$.}
 \begin{center}
\begin{tabular}{ccc cc ccc c c c c}
\hline \hline
\multicolumn{3}{c}{Redshift} & & Area &N& \multicolumn{2}{c}{ $\log(L$ [\ulum])}& Type & Source\\
mean & min & max & ${\rm\log(\frac{V}{Mpc^3})}$ & (deg$^2$) && min & max \\ \hline
\hline
0.36& 0.33& 0.40& 4.97& 0.60& 326 & 40.5& 42.8 & S & DEEP2 \\
0.55& 0.50& 0.60& 5.39& 0.60& 211 & 40.5& 42.8 & S & - \\
0.65& 0.60& 0.70& 5.48& 0.60& 319 & 41.0& 42.8 & S & - \\
0.74& 0.70& 0.78& 6.12& 2.78& 550 & 41.1& 42.8 & S & - \\
0.54& 0.41& 0.65& 5.75& 0.61& 325 & 40.8& 42.2 & S & VVDS-Deep \\ \hline
0.41 & 0.39 & 0.41 & 4.09 & 0.24& 2219 & 39.5 & 41.3 &NB & \citet{Ly_2007}\\
0.50 & 0.10 & 0.90 & 5.79 & 0.20 & 401 & 39.5 & 41.5 &S & \citet{Pirzkal2013} \\
0.83 & & & 5.09 & 0.38& 910 & 41.0 & 41.5 &NB & \citet{Drake_2013} \\
0.99 & & & 5.17 & 0.38& 32 & 41.5 & 42.0 &NB & - \\
1.15 & 0.7 & 1.5 & 6.84 & 1.04 & 155 & 41.7 & 42.7 &S & \citet{Colbert2013} \\
1.85 & 1.5 & 2.2 & 6.85 & 1.04 & 54 & 42.2 & 43.0 &S & - \\
\hline
\hline
\end{tabular}
\end{center}
\label{OIII:LFs:literature}
\end{table*}

\subsection{LF model}

To investigate the LFs and their evolution more quantitatively, we explore two analytic models: the \citet{Schechter1976} and \citet{Saunders_1990} models, 
both with parameters explicitly dependent on redshift.

Like broad-band galaxy LFs, the emission line LF is often modelled with 
a three-parameter \citet{Schechter1976} model:
\begin{equation}
\Phi(L){\rm d}L=\phi_* \left( \frac{L}{L_*}\right)^\alpha \exp{\left(-\frac{L}{L_*}\right)} \, {\rm d}\, \left(\frac{L}{L_*}\right)\,\mathrm{,}
\label{eqn:schechter}
\end{equation}
with $\Phi_*$, $L_*$ and $\alpha$ its parameters representing the density and the luminosity of typical ELGs and the faint-end slope.
Recent investigations have found that the \OII emission line LF when sampled at its brightest end is better represented by a double power-law form \citep{Zhu_2009,Gilbank_2010,comparat_2015_OII}, which declines less steeply than an exponential as in the Schechter model. This is likely because the Schechter model is most suited for characterizing the stellar mass function, while line emission originates from star formation and the line LF, to the first order, must share the same shape as the SFR function. Recent observations have shown that the SFR function at the high end declines less fast than the Schechter function \citep[][ and references therein]{Salim_2012}.
We fit our measurements with a four-parameter \citet{Saunders_1990} model, which is a re-parametrization of a double power-law:
\begin{equation}
\Phi(L){\rm d}L = \phi_* \left(\frac{L}{L_*}\right)^{\alpha} \exp{\left[ -\left(\frac{\log_{10}(1+L/L_*)}{\sqrt{2}\sigma}\right)^2\right]} {\rm d}L \,\mathrm{.}
\label{eqn:saunders}
\end{equation}
with $\Phi_*$, $L_*$, $\alpha$ and $\sigma$ its parameters representing the density, the luminosity of typical ELGs, the faint-end slope and the width of the transition between the bright and the faint end. Note that these parameters do not have the same meaning with the two models.

For both models, we allow linear redshift dependence for the parameters $L_*$, $\Phi_*$ and $\alpha$:
$L_*(z) = L_*(0)(1+z)^{\beta_L}$, $\Phi_*(z)=\Phi^*(0)(1+z)^{\beta_\Phi}$, $\alpha(z)=\alpha(0)(1+z)^{\beta_\alpha}$
and fit all the measurements over the entire redshift range simultaneously. 
In all data sets, there is no need to parametrize the redshift dependence of $\alpha$ (data is not sufficient to constrain its eventual evolution).
In the Saunders model, we find a transition parameter $\sigma=0.54\pm0.2$ fits the \OII data well and allows for a smoother transition between the faint and the bright populations. Though this parameter is not well constrained, so in the final analysis, we fix the value $\sigma=0.54$ for all Saunders fits.

\begin{figure*}
\begin{center}
\includegraphics[width=16cm]{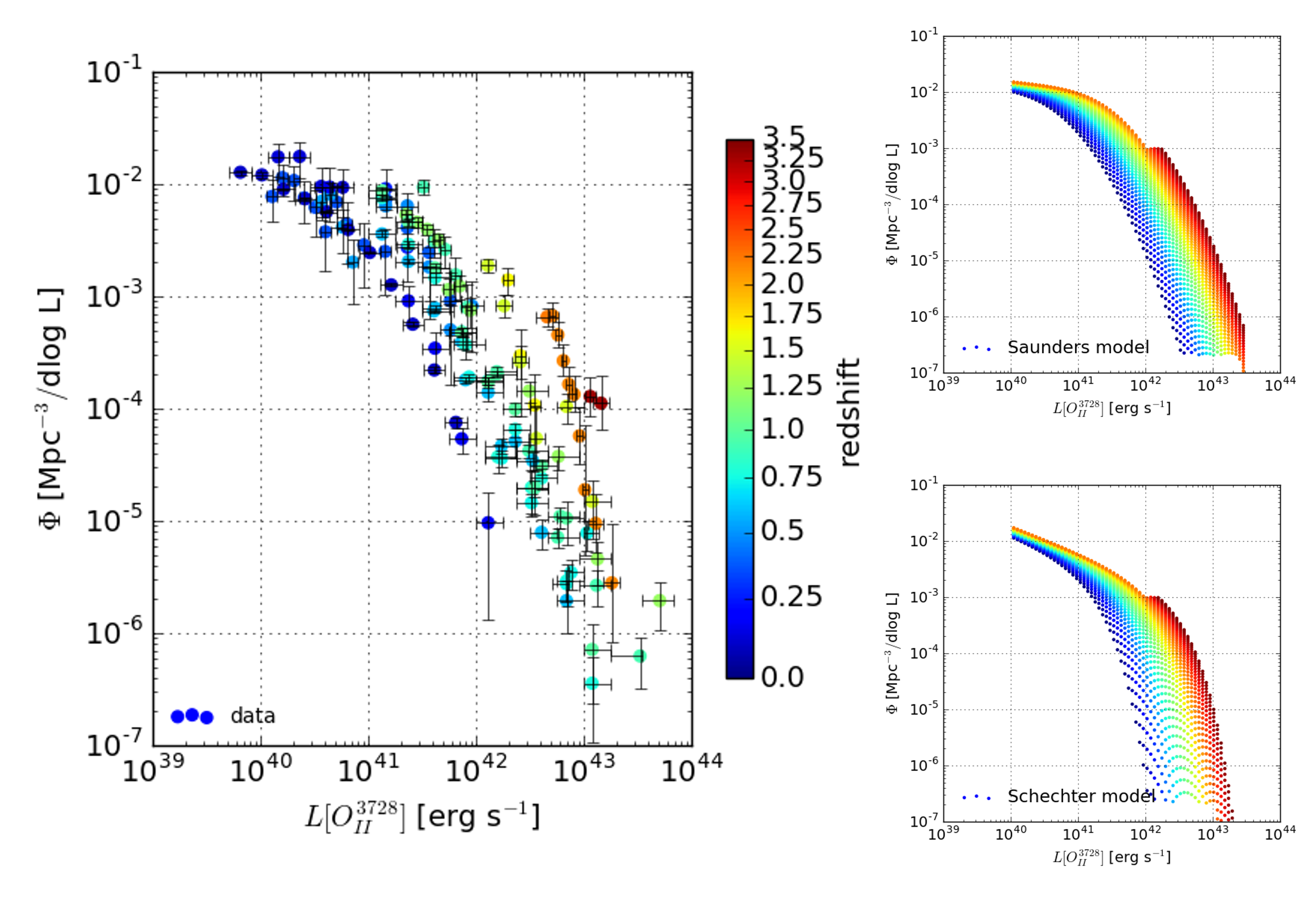}
\caption{Left: the observed \OIIll LF. Right: the best-fitting Schechter and Saunder model models. Error bars come from jackknife re-sampling for our measurements. The redshift colour coding is the same for all panels.}
\label{fig:modelled:LF:O2}
\end{center}
\end{figure*}

\begin{figure*}
\begin{center}
\includegraphics[width=16cm]{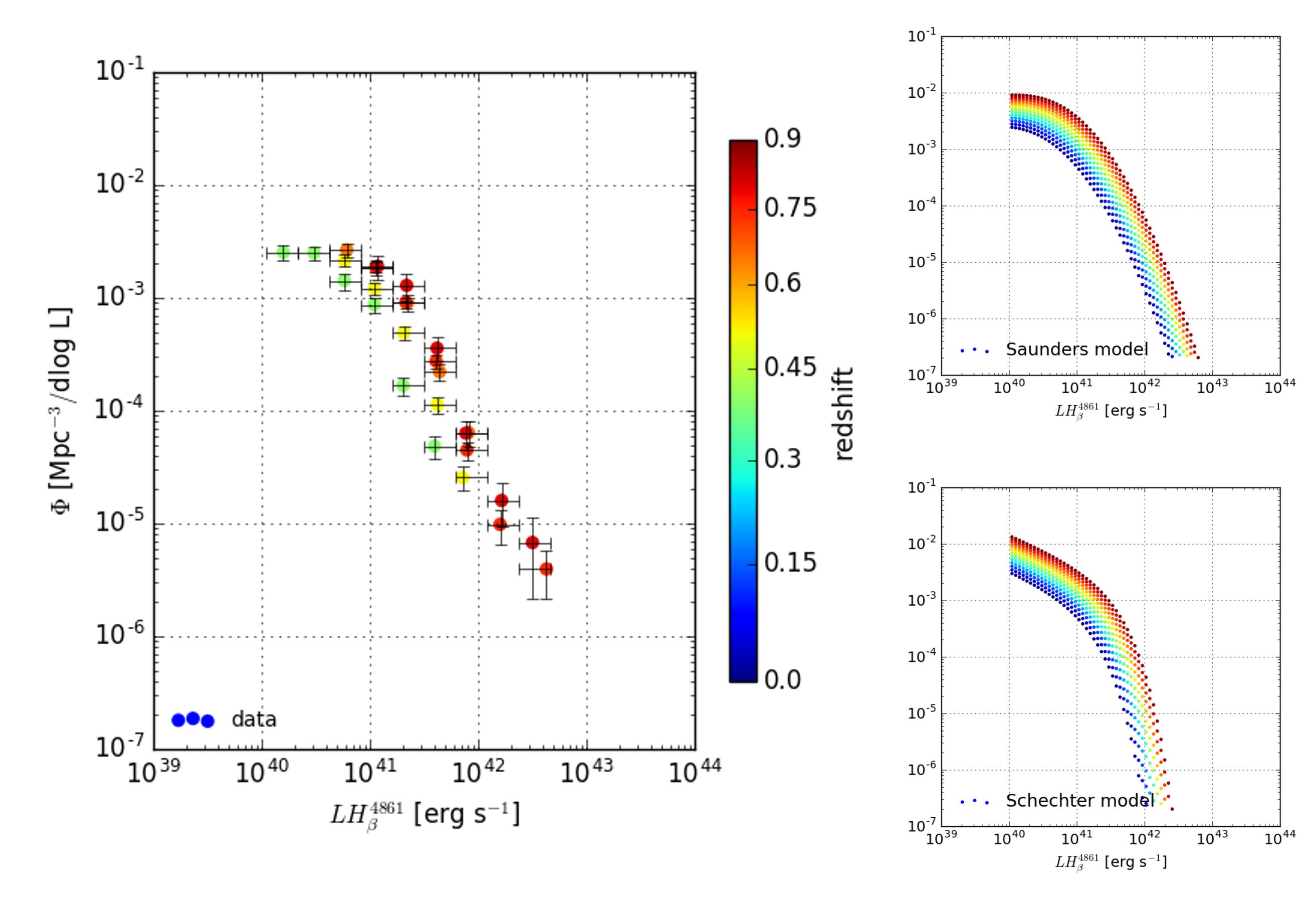}
\caption{Same as Fig. \ref{fig:modelled:LF:O2} for the \Hbl LF.}
\label{fig:modelled:LF:Hb}
\end{center}
\end{figure*}

\begin{figure*}
\begin{center}
\includegraphics[width=16cm]{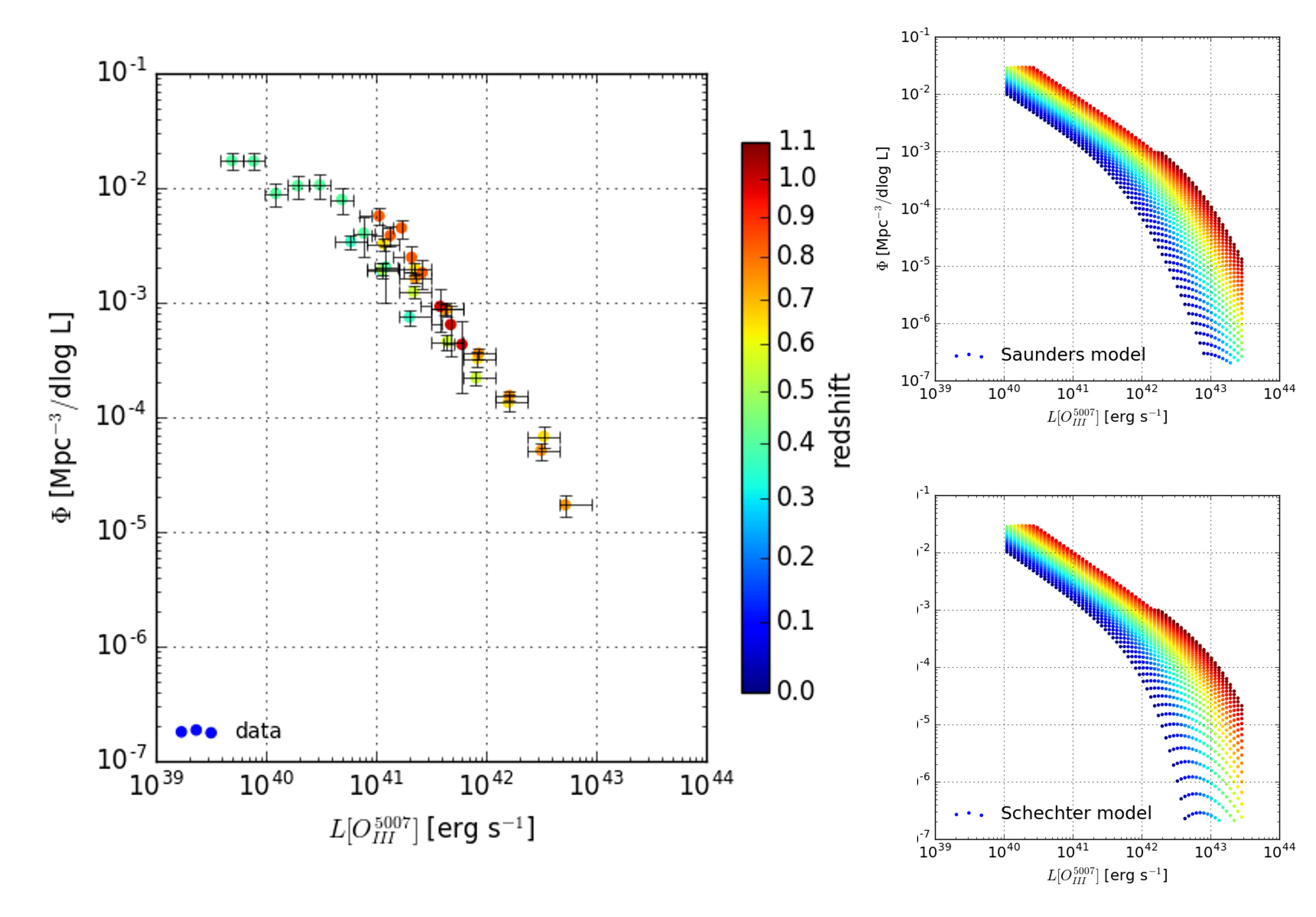}
\caption{Same as Fig. \ref{fig:modelled:LF:O2} for the \OIIIb LF.}
\label{fig:modelled:LF:O3}
\end{center}
\end{figure*}

We present the best-fitting models in Table~\ref{table:LFs:fits} and Figs.~\ref{fig:modelled:LF:O2} (for \OII), \ref{fig:modelled:LF:Hb} (for \Hb) and \ref{fig:modelled:LF:O3} (for \OIII). 
For all the lines, both $L_*$ and $\Phi_*$ increase with redshift, indicating brighter average luminosity and more strong emitters at higher redshift.
We also find that for the \Hb LF, the faint-end slope is similar to that of the \Ha LF from the literature \citep[e.g.,][]{2013MNRAS.428.1128S}.
Finally, although we find that both models account well for the \Hb and \OIII data, the Saunders model gives slightly better values of $\chi^2$. For the \OII LF, the best Schechter model yields a reduced $\chi^2=2.77$ when the best Saunders model has 1.77. We join the conclusions of \citet{Zhu_2009} that showed a double power-law accounted well for the observed \OII LF: the bright end decline of an exponential seems too sharp to model the data.

The global fit presented here is in agreement with former individual results by construction, indeed the data from previous studies is included for the fit. In the literature, Schechter functions are typically fitted on a single redshift bin and on a single galaxy sample. Therefore it is complicated to present a face to face value comparison between each parameter obtained as faint-end slopes, and to a lesser extent cosmological parameters, may vary from a paper to another. We face the same issue when comparing to double power-laws. At redshift $z=0$, \citet{Gilbank_2010} fitted the \OII LF ($L_*$, $\alpha$)=($10^{41.3}$\ulum,-1.3) and we find ($10^{41.1}$\ulum, -1.4), which is very close. 
\citet{Ly_2007} fitted the \OIII LF with a similar faint-end slope as us and they obtain $(\log[L_*(z=0.42)($\ulum$)], \log[L_*(z=0.83)($\ulum$)]) = (41.7\pm0.4, 42.2\pm0.1)$ where we have $(42.0\pm0.1, 42.4\pm0.1)$.

Based on the Saunders models, we compute the expected number of sources per deg$^2$ at a given redshift and luminosity. We show this prediction for a density of 1 deg$^{-2}$ as a function of redshift and luminosity for the three lines along with the observed density of data on Fig. \ref{fig1:redshift:luminosity} (top solid line). Since the surveys used in this analysis are of order of a square degree, this solid line constitutes the bright luminosity limit these surveys can probe within the volume they sample.

\begin{table*}
\caption{The best-fitting Schechter and Saunders redshift dependent models. For each line and each model, we give the values of the parameters characterizing equations (\ref{eqn:schechter}) and (\ref{eqn:saunders}). For Saunders model's fits we fix $\sigma$ to 0.54. Models are illustrated in Figs.~\ref{fig:modelled:LF:O2} (for \OII), \ref{fig:modelled:LF:Hb} (for \Hb) and \ref{fig:modelled:LF:O3} (for \OIII).}
\begin{center}
\begin{tabular}{ccccccccccccccccc}
\hline \hline
line & model &$\chi^2$/d.o.f & \multicolumn{2}{c}{$L_*(z)$ [\ulum] $= L_*(0)(1+z)^{\beta_L}$} & \multicolumn{2}{c}{$\Phi_*(z) [$Mpc$^{-3}] =\Phi^*(0)(1+z)^{\beta_\Phi}$} & \multicolumn{2}{c}{$\alpha(z)=\alpha(0)$} \\ 
 & & & $\log_{10}(L^*(0))$ & $\beta_L$ & $\log_{10}(\Phi^*(0))$& $\beta_\phi$ & $\log_{10}(\alpha(0))$ \\ \hline

\OII & Schechter&
2.77& 
41.1$^{0.02}_{-0.03}$ & 
2.33$^{0.14}_{-0.16}$ & 
-2.4$^{0.03}_{-0.03}$ & 
-0.73$^{0.25}_{-0.29}$ & 
-1.46$^{0.06}_{-0.05}$ \\
\OII& Saunders &
1.77& 
40.1$^{0.02}_{-0.03}$ & 
1.92$^{0.13}_{-0.16}$ & 
-1.95$^{0.03}_{-0.03}$ & 
0.07$^{0.24}_{-0.28}$ & 
-1.12$^{0.04}_{-0.04}$ 
 \\

 \hline

\Hb & Schechter &
0.62& 
40.88$^{0.05}_{-0.07}$ & 
2.19$^{0.25}_{-0.32}$ & 
-3.34$^{0.09}_{-0.12}$ & 
2.7$^{0.44}_{-0.57}$ & 
-1.51$^{0.27}_{-0.2}$ \\

\Hb & Saunders & 
0.45& 
39.7$^{0.06}_{-0.07}$ & 
1.63$^{0.26}_{-0.34}$ & 
-2.92$^{0.09}_{-0.11}$ & 
3.37$^{0.44}_{-0.56}$ & 
-0.81$^{0.07}_{-0.1}$ \\

 \hline

\OIII& Schechter & 
0.61& 
41.42$^{0.07}_{-0.09}$ & 
3.91$^{0.32}_{-0.4}$ & 
-3.41$^{0.08}_{-0.1}$ & 
-0.76$^{0.39}_{-0.49}$ & 
-1.83$^{0.1}_{-0.08}$ \\

\OIII& Saunders &
0.56& 
40.81$^{0.07}_{-0.09}$ & 
3.31$^{0.32}_{-0.4}$ & 
-2.91$^{0.08}_{-0.1}$ & 
-0.22$^{0.39}_{-0.49}$ & 
-1.81$^{0.17}_{-0.13}$ \\

\hline
\hline
\end{tabular}
\end{center}
\label{table:LFs:fits}
\end{table*}

\section{Summary}

We have collected the spectroscopic data from two deep surveys (VVDS, DEEP2) and 
measured the LFs of three emission lines, \OIIll, \Hbl and \OIIIb
at moderate redshifts ($0.2 \lesssim z \lesssim 1.3$) using 35,639 galaxy spectra.

We compiled previous measurements from the literature and performed analytic fitting to the entire data sets of each emission line with both \citet{Schechter1976} and \citet{Saunders_1990} models, allowing natural redshift dependence of the parameters $L^*$ and $\phi^*$. For all lines, we find previous measurements to be compatible with newer measurements and they can be modelled together with a natural redshift evolution. This compilation of the literature along with the new measurements reaches large volumes: $\sim5\times10^7$ Mpc$^3$ for \OII, $\sim3\times10^6$ Mpc$^3$ for \Hb, $\sim2\times10^7$ Mpc$^3$ for \OIIIb. 

We find that for all the three lines, the characteristic luminosity and density increase with redshift. Using the Schechter model over the redshift ranges considered, we find that, for \OII emitters, the characteristic luminosity $L_*(z=0.5)=3.2\times10^{41}$\ulum increases by a factor of $2.7 \pm 0.2$ from z=0.5 to z=1.3, for \Hb emitters $L_*(z=0.3)=1.3\times10^{41}$\ulum by a factor of $2.0 \pm 0.2$ from z=0.3 to z=0.8, and for \OIII emitters $L_*(z=0.3)=7.3\times10^{41}$\ulum by a factor of $3.5 \pm 0.4$ from z=0.3 to z=0.8. It indicates that on average, the emitters are more numerous and luminous at higher redshift.

This measurement is crucial for the development of truthful ELGs mock catalogues based on $N$-body simulations and semi-analytical models \citep[e.g.][]{galform_GP2014,orsi14}. In the future, we hope to compare such mock catalogues with semi-analytical models of galaxy formation with these measurements and hereby obtain a better understanding of the links between line luminosity, SFR and dust in ELGs at redshift one. In this aim, we will study in the near future the conditional emission line LFs and line ratios distributions. On the longer term, we aim to understand precisely the place of this ELG population within the global paradigm of galaxy formation and evolution.
This is utmost important to make sure the planned ELG-BAO measurements will not be affected by systematics due to selection effect.

Finally, our data and measurements is made publicly available and we provide a \textsc{python} package to mine further the information available in this data set. The framework developed is flexible so that any new data set can be seamlessly folded into the luminosity function fits.

\bibliographystyle{mn2e}
\bibliography{biblio}

\section*{Acknowledgements}
\vspace{0.2cm}

JC and FP acknowledge support from the Spanish MICINNs Consolider-Ingenio 2010 Programme under grant MultiDark CSD2009-00064, MINECO Centro de Excelencia Severo Ochoa Programme under the grants SEV-2012-0249, FPA2012-34694, and the projects AYA2014-60641-C2-1-P and AYA2012-31101. GY acknowledges financial support from MINECO (Spain) under project number AYA2012-31101 and AYA2015-63810.
PN acknowledges the support of the Royal Society through the award
of a University Research Fellowship, the European Research
Council, through receipt of a Starting Grant (DEGAS-259586).
and the support of the Science and Technology Facilities Council
(ST/L00075X/1).
TD and JPK acknowledge support from the LIDA ERC advanced grant. AR acknowledges funding from the P2IO LabEx (ANR-10-LABX-0038) in the framework Investissements d'Avenir (ANR- 11-IDEX-0003-01) managed by the French National Research Agency (ANR). EJ and LT acknowledge the support of CNRS and the Labex OCEVU. This work has been carried out thanks to the support of the OCEVU Labex (ANR-11-LABX-0060) and the A*MIDEX project (ANR-11-IDEX-0001-02) funded by the "Investissements d'Avenir" French government programme managed by the ANR.

\end{document}